\documentclass[12pt,epsf]{article}
\usepackage{epsfig}

\newcommand{\setfigure}[2]{\begin{figure}[htbp]
\begin{center}\leavevmode\epsfxsize=5in\epsfbox{#1.eps}\end{center}\caption{#2\label{#1}}
\end{figure}}

\renewcommand{\thanks}[1]{\footnote{#1}} 

\newcommand{\be}{\begin{equation}}
\newcommand{\ee}{\end{equation}}
\newcommand{\bea}{\begin{eqnarray}}
\newcommand{\eea}{\end{eqnarray}}

\begin{document}

\pagestyle{empty}

\bigskip\bigskip
\begin{center}
{\bf \large Radial Photon Trajectories Near an Evaporating Black Hole}
\end{center}

\begin{center}
Beth A. Brown\footnote{e-mail address, beth.a.brown@nasa.gov}\\
James Lindesay\footnote{e-mail address, jlindesay@fac.howard.edu} \\
Computational Physics Laboratory \\
Howard University,
Washington, D.C. 20059 
\end{center}
\bigskip

\begin{center}
{\bf Abstract}
\end{center}
The radial motion of photons emitted near the
horizon of a black hole that evaporates at a
steady rate is examined.  The space-time of the
black hole is generated using non-orthogonal coordinates. 
It is shown that some
photons that are initially drawn towards the singularity
can escape falling into the horizon.  The behaviors
of various outgoing and ingoing photons are clearly
demonstrated through the use of a Penrose diagram.
\bigskip \bigskip \bigskip

\setcounter{equation}{0}
\section{Introduction}
\indent

A black hole is an object with an extreme gravitational field
whose escape velocity exceeds the speed of light. Classical treatments of static,
spherically symmetric black holes generally use the Schwarzschild solution to
Einstein's equations of general relativity.  For astrophysical black holes which
likely undergo periods of accretion and evaporation,
it is more realistic to consider dynamic rather than static behavior. 

We have examined the geometry of an example dynamic black hole that
evaporates at a steady rate, as measured by a distant observer.  
A flat Minkowski space-time results once the mass
of the black hole vanishes\cite{JLOct07}. 

The metric defining that space-time considered is given by 
\be
ds^2 = -\left (1-{R_M (ct) \over r} \right ) (dct) ^2 +
2 \sqrt{{R_M (ct) \over r}} \, dct \, \, dr + dr^2 + r^2 \, d \omega ^2
\label{metric}
\ee
where $d\omega^2 \equiv d\theta^2 + sin^2 \theta \, d\phi^2$,
and it is assumed\cite{JLMay07} that $\ddot{R}_M =0$.   
In this equation, $R_M \equiv 2 G_N M(ct) / c^2$ is a dynamic
form of the Schwarzschild radius that we refer to as the
\emph{radial mass scale}.  The metric in Eq. \ref{metric} is
seen to have anomalous behavior at $r=R_M (ct)$.

Photons are given by null geodesics. The  radial null surfaces ($ds^2 =0$)
for the metric specified in Eq. \ref{metric} satisfy
\be
{d r_\gamma \over dct} = - \sqrt{R_M \over r_\gamma} \pm 1,
\label{photons}
\ee
where the +/- sign describes outgoing/ingoing photons. 
The radial coordinate corresponding to the horizon satisfies
\be
{d R_H \over dct} = - \sqrt{R_M \over R_H} + 1 \quad , \quad 
R_H = {R_M \over \left (1 - \dot{R}_H  \right ) ^2} < R_M \quad ,
\label{horizon}
\ee
and is given by
the particular null surface proportional to the radial mass scale
$R_H=R_M / \zeta_H$, where
\be
\dot{R}_M = 
\zeta_H \, \left(1-\sqrt{\zeta_H} \right ) .
\label{zetaH}
\ee
The ability to relate the horizon scale to the radial mass scale
in terms of a proportionality constant $\zeta_H$
allows this algebraic solution for a given rate of
evaporation.

We examine the radial motion of photons emitted near the black hole horizon. 
It is shown that some photons emitted inside 
the radial mass scale (or ``shrinking Schwarzschild radius")
can escape falling into the horizon, and that this behavior is clearly demonstrated
through the use of the Penrose diagram.

\setcounter{equation}{0}
\section{Behavior of Near-Horizon Photons
\label{mainsection}}

\subsection{Conformal coordinates for Penrose diagram}
\indent

The form of the conformal time and radial coordinates
that are used for the construction of the Penrose diagram
have been developed in a companion paper\cite{JLOct07}:
\be
\begin{array}{c}
ct_* = {r \over 2} \left (
 exp \left [ \int ^ {R_M (ct) \over r} { \left (
1 + \sqrt{\zeta '}  \right ) d \zeta '  \over
\left \{ \zeta' \left ( 1 + \sqrt{\zeta'} \right )
+ \dot{R}_M \right \} }
\right ]  -  
 exp \left [ \int ^ {R_M (ct) \over r} { \left (
1 - \sqrt{\zeta '}  \right ) d \zeta '  \over
\left \{ \zeta' \left ( 1 - \sqrt{\zeta'} \right )
- \dot{R}_M \right \} }
\right ] \right ) \\ \\
r_* = {r \over 2} \left (
 exp \left [ \int ^ {R_M (ct) \over r} { \left (
1 + \sqrt{\zeta '}  \right ) d \zeta '  \over
\left \{ \zeta' \left ( 1 + \sqrt{\zeta'} \right )
+ \dot{R}_M \right \} }
\right ]  + 
 exp \left [ \int ^ {R_M (ct) \over r} { \left (
1 - \sqrt{\zeta '}  \right ) d \zeta '  \over
\left \{ \zeta' \left ( 1 - \sqrt{\zeta'} \right )
- \dot{R}_M \right \} }
\right ] \right )  .
\end{array}
\label{conformal}
\ee
This equation relates the space-time coordinates of an
asymptotic observer $(ct, r)$ with the conformal coordinates
$(ct_* , r_*)$.

\subsection{Outgoing photon trajectories}
\indent

We'll first examine the trajectories of outgoing photons corresponding to
the + sign in Eqn \ref{photons}. 
All photons are emitted at $ct = -10$ units from various
initial radial coordinates.  The calculated trajectories 
are shown in space-time plots in Fig. \ref{Outgo}.
\setfigure{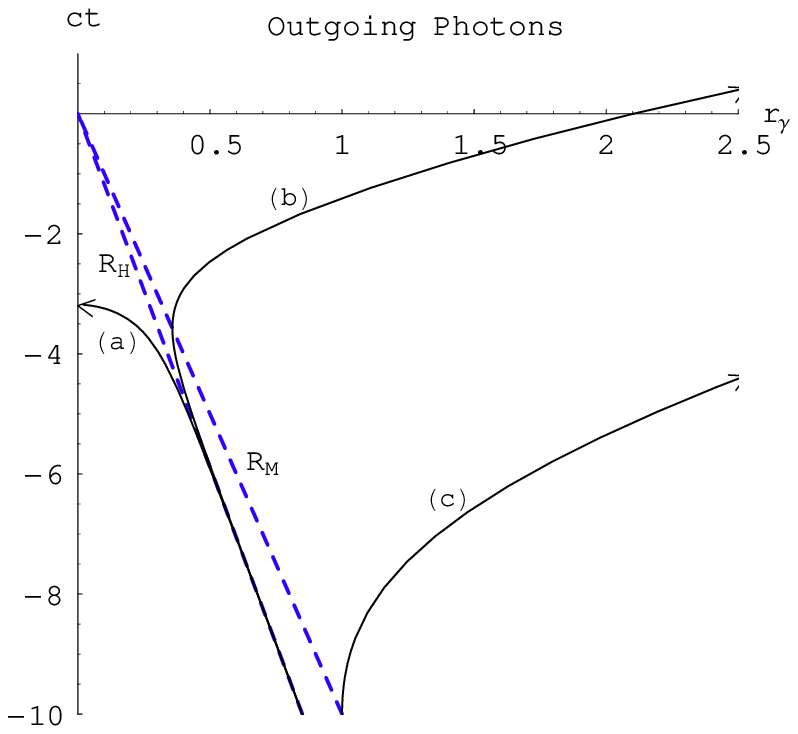}{Trajectories of outgoing photons emitted
(a) inside $R_H$, (b) between $R_H$ and $R_M$, and
(c) at $R_M$.}
In this figure, the dashed line closest to the singularity
represents the horizon of the black hole $R_H (ct)$,
while the outer dashed line represents the radial
mass scale $R_M (ct)$ (the coordinate anomaly).
The photon labeled (a) starts inside the black hole horizon. 
Although it is an outgoing photon, 
the radial coordinate of the photon $r_\gamma$ decreases monotonically
until it reaches the singularity at $r=0$. 

The photon labeled (b) starts between the horizon $R_H$ and the
radial mass scale $R_M$. 
Initially, the photon falls in toward the singularity ($r =0$),
tracking (but remaining just outside) the horizon.  
This outgoing photon falls inward due to the strong gravity near the black hole. 
In the late stages of evaporation, 
gravity weakens enough to allow the photon to move away from the horizon. 
As $r_\gamma$ crosses
$R_M$, the photon is momentarily stationary in the $(ct,r)$ coordinates. 
The slope of the photon
approaches 45$^\textnormal{\tiny{o}}$ at spatial infinity as it
escapes the influence of the black hole.

The photon labeled (c) starts on the coordinate anomaly, $R_M$ .  
The photon is initially stationary at $R_M$,
and at no point does $r_\gamma$ decrease. 
The coordinate anomaly shrinks away from the photon as the black hole evaporates
and gravity weakens, and the photon subsequently escapes.

The trajectories of these outgoing photons
are illustrated on the Penrose diagram in Fig. \ref{OutPen}.  
\setfigure{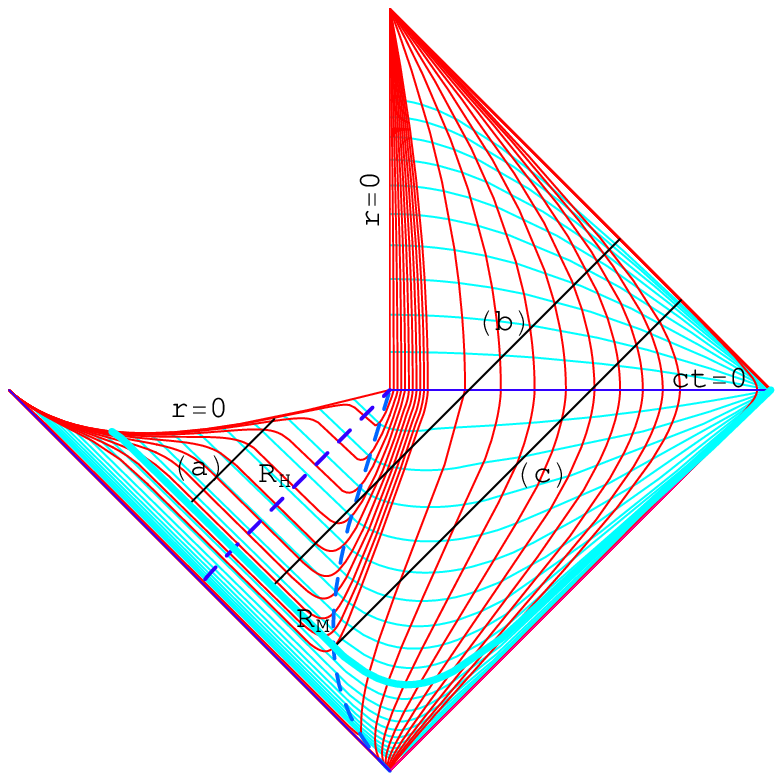}{Penrose diagram of an evaporating black
hole with outgoing photon trajectories over plotted.  Red curves
(running vertically in right hand region) represent curves of
constant $r$.  The blue-green curves represent curves of constant $ct$.}
In this Penrose diagram, the space-time structure is represented using
functions of the conformal coordinates Eq. \ref{conformal}
developed in the companion paper \cite{JLOct07}. 
On the Penrose diagram, the path of photons are immediately apparent,
since all light-like surfaces lie at 45$^\textnormal{\tiny{o}}$ angles.

\subsection{Ingoing photon trajectories}
\indent

We will next examine the trajectories of ingoing photons. 
For an evaporating black hole, there exists a (ingoing) \emph{light-like
boundary} that separates the regions of space-time for
which any object can communicate with the singularity. 
This shrinking boundary, with radial coordinate
$R_B$, is shown as an additional dashed
line outside of the horizon and radial mass scale in the
space-time plot of Fig. \ref{Ingo}. 
\setfigure{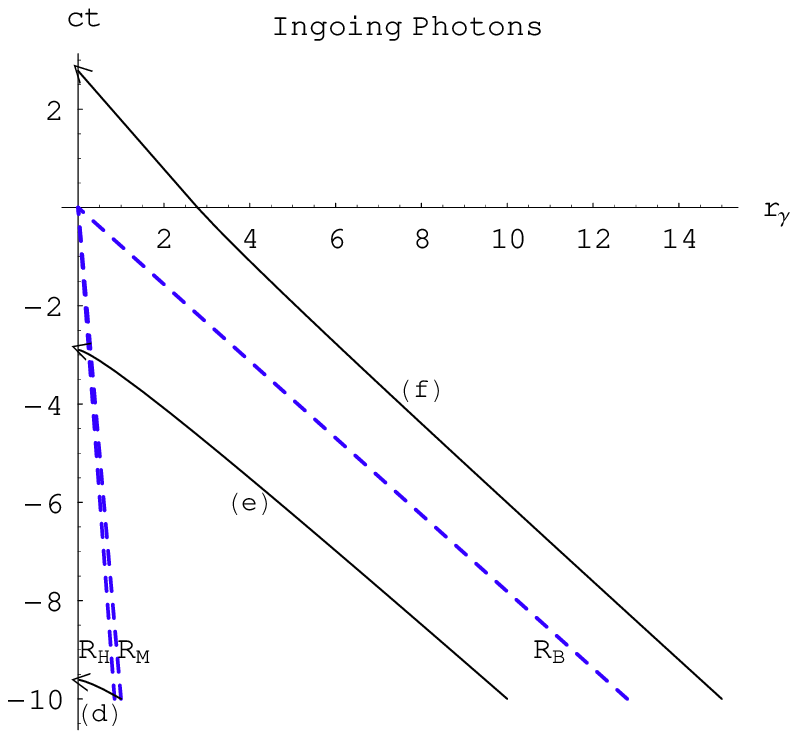}{Trajectories of ingoing photons emitted
(d) at $R_M$, (e) between $R_M$ and $R_B$,
and (f) outside $R_B$.}
As with the outgoing photons, all ingoing photons are
emitted at ct=-10 units. 
The photon labeled (d) starts on the coordinate anomaly,
crosses the horizon, and hits the singularity. 
This photon reaches the singularity within an extremely short time interval after emission.
The photon labeled (e) starts between the coordinate anomaly
$R_M$ and the shrinking light-like boundary $R_B$.  
The photon crosses the coordinate anomaly $R_M$ and the black hole event horizon
$R_H$  (the outgoing causal horizon), 
and hits the singularity.
The photon labeled (f) starts outside the shrinking
light-like boundary $R_B$, and never crosses the black hole horizon. 
These trajectories are plotted on the Penrose diagram in Fig. \ref{InPen}.
\setfigure{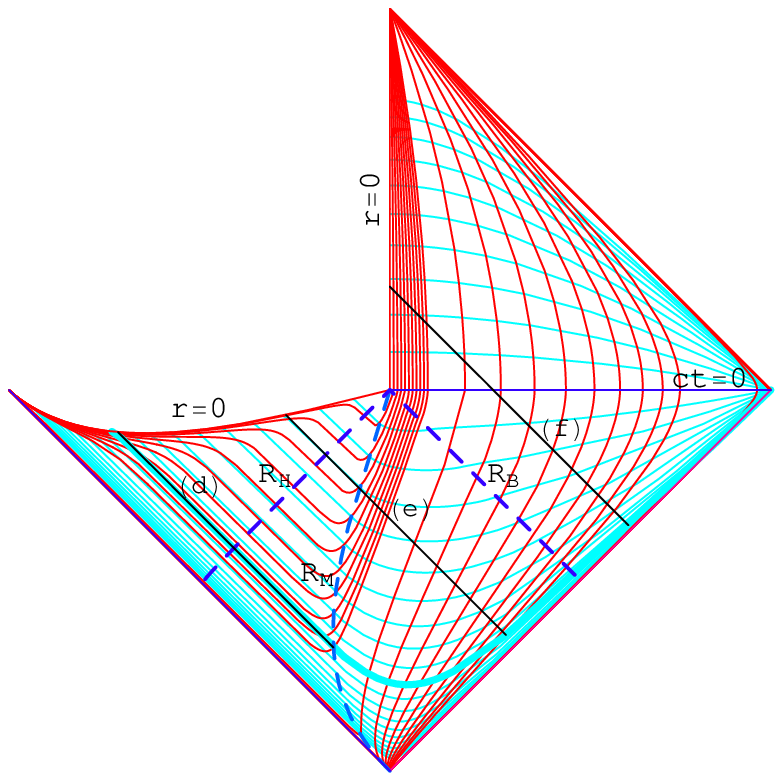}{Penrose diagram with ingoing photon trajectories over plotted.}

\setcounter{equation}{0}
\section{Conclusions}
\indent

We have calculated the paths of outgoing and ingoing photons
originating near an evaporating black hole.  Several behaviors
are found to be of interest.  It is found that outgoing photons
just outside the horizon initially approach the singularity, until the
mass of the black hole decreases to a point that allows the
photons to alter direction and begin to move away from the
singularity.

When plotted on the Penrose diagram for this space-time,
the photon trajectories confirm the expected conformal
coordinate representation.  Consistent with expectations, outgoing photons are always
stationary at $r=R_M$; it is the shrinking mass scale $R_M$
that moves away from those photons.  In addition, because the black
hole eventually evaporates away, there is a class of
photons that can never reach the singularity.

Analogous results have been obtained for an accreting black
hole, which will be presented in a subsequent paper.

\begin{center}
\textbf{Acknowledgments}
\end{center}
BAB would like to acknowledge the support of the NASA Administrator's
Fellowship Program.

\end{document}